\documentclass[
superscriptaddress,aip,apl,reprint,twocolumn,amsmath,amssymb,showpacs]{revtex4-2}
\usepackage{graphicx}  
\usepackage{mathptmx}
\usepackage{epstopdf}
\usepackage{dcolumn}   
\usepackage{bm}        
\usepackage{amssymb}   
\usepackage{amsmath}   
\usepackage{amsthm}
\usepackage{chemarrow}
\usepackage{color}
\usepackage{mathrsfs}
\usepackage{float}
\usepackage{bbold}
\usepackage{bbm}
\usepackage{braket}
\usepackage{physics}
\usepackage{titlesec} 
\usepackage[a4paper,colorlinks=true,
linkcolor=blue,citecolor=blue,
pdfauthor={ },
pdftitle={ },
pdfsubject={ },
pdfkeywords={ }]{hyperref}

\theoremstyle{plain}
\newtheorem*{theorem*}{Theorem}

\titleformat{\section}{\large\sffamily\bfseries}{\thesection}{}{}
\titleformat{\subsection}[runin]{\sffamily\bfseries}{\thesubsection}{}{}
\titlespacing*{\section}{0pt}{3ex}{0ex}
\titlespacing*{\subsection}{0pt}{2ex}{1ex}

\usepackage{verbatim}

\begin{document}


\title{Enhanced quantum hypothesis testing via the interplay between coherent evolution and noises}

%

\author{Qing Li}
\thanks{These authors contributed equally.}
\affiliation{
CAS Key Laboratory of Microscale Magnetic Resonance and School of Physical Sciences, University of Science and Technology of China, Hefei, Anhui 230026, China}
\affiliation{
CAS Center for Excellence in Quantum Information and Quantum Physics, University of Science and Technology of China, Hefei, Anhui 230026, China}
\affiliation{
\mbox{Hefei National Laboratory, University of Science and Technology of China, Hefei 230088, China}}

\author{Lingna Wang}
\thanks{These authors contributed equally.}
\affiliation{
Department of Mechanical and Automation Engineering, The Chinese University of Hong Kong}

\author{Min Jiang}
\email[]{dxjm@ustc.edu.cn}
\affiliation{
CAS Key Laboratory of Microscale Magnetic Resonance and School of Physical Sciences, University of Science and Technology of China, Hefei, Anhui 230026, China}
\affiliation{
CAS Center for Excellence in Quantum Information and Quantum Physics, University of Science and Technology of China, Hefei, Anhui 230026, China}
\affiliation{
\mbox{Hefei National Laboratory, University of Science and Technology of China, Hefei 230088, China}}

\author{Ze Wu}
\affiliation{
CAS Key Laboratory of Microscale Magnetic Resonance and School of Physical Sciences, University of Science and Technology of China, Hefei, Anhui 230026, China}
\affiliation{
CAS Center for Excellence in Quantum Information and Quantum Physics, University of Science and Technology of China, Hefei, Anhui 230026, China}
\affiliation{
\mbox{Hefei National Laboratory, University of Science and Technology of China, Hefei 230088, China}}

\author{Haidong Yuan}
\email[]{hdyuan@mae.cuhk.edu.hk}
\affiliation{
Department of Mechanical and Automation Engineering, The Chinese University of Hong Kong}

\author{Xinhua Peng}
\email[]{xhpeng@ustc.edu.cn}
\affiliation{
CAS Key Laboratory of Microscale Magnetic Resonance and School of Physical Sciences, University of Science and Technology of China, Hefei, Anhui 230026, China}
\affiliation{
CAS Center for Excellence in Quantum Information and Quantum Physics, University of Science and Technology of China, Hefei, Anhui 230026, China}
\affiliation{
\mbox{Hefei National Laboratory, University of Science and Technology of China, Hefei 230088, China}}

\date{\today}

\begin{abstract}
Previous studies in quantum information have recognized that specific types of noise can encode information in certain applications. However, the role of noise in Quantum Hypothesis Testing (QHT), traditionally assumed to undermine performance and reduce success probability, has not been thoroughly explored. Our study bridges this gap by establishing sufficient conditions for noisy dynamics that can surpass the success probabilities achievable under noiseless (unitary) dynamics within certain time intervals.
We then devise and experimentally implement a noise-assisted QHT protocol in the setting of ultralow-field nuclear magnetic resonance spin systems. Our experimental results demonstrate that the success probability of QHT under the noisy dynamics can indeed surpass the ceiling set by unitary evolution alone. Moreover, we have shown that in cases where noise initially hampers the performance, strategic application of coherent controls on the system can transform these previously detrimental noises into advantageous factors. This transformative approach demonstrates the potential to harness and leverage noise in QHT, which pushes the boundaries of QHT and general quantum information processing.

\end{abstract}

\maketitle

Hypothesis testing is an essential statistical technology in scientific research,
enabling one to distinguish various models based on observed data\,\cite{RevModPhys.71.S96}.  
In quantum science,
quantum hypothesis testing (QHT) is a commonly employed tool to ascertain the model of a given quantum system,
which has profound connections with topics ranging from quantum state and dynamics discrimination\,\cite{holevo1973statistical, RN6, RN14, RN15, RN16, RN39}, parameter estimation\,\cite{ RN22, RN23}, quantum communication\,\cite{RN26,wang2012one}, the detection of weak forces and magnetic fields\,\cite{RN9, RN10, RN12}, quantum machine learning\,\cite{weber2021optimal}, to quantum illumination\,\cite{wilde2017gaussian}.

In noiseless scenarios, it is possible to achieve perfect hypothesis testing within a sufficient amount of time using optimal strategies\cite{RN27, RN28}. However, the presence of inherent quantum noise significantly hampers this capability. To mitigate the impact of noise, various strategies, such as quantum error correction \cite{zhou2018achieving,dur2014improved, PhysRevA.100.022336} and optimal control \cite{PhysRevResearch.2.033396}, have been proposed. 
Although recent studies show that some specific types of noise can be harnessed in certain quantum information applications\cite{kukita2021heisenberg}, it is widely accepted that the performance of noisy hypothesis testing is fundamentally limited by its noiseless limit. The potential to leverage noise in QHT is still unexplored, particularly beyond specific scenarios.

\begin{figure}  
	\makeatletter
	\def\@captype{figure}
	\makeatother
	\includegraphics[scale=1]{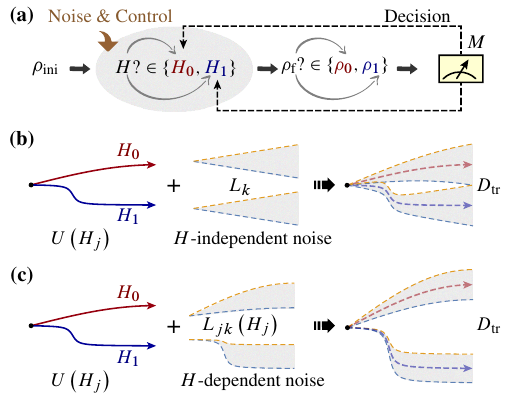} %
	\caption{\textbf{Exemplary illustration of quantum hypothesis testing.} (a) Binary hypothesis testing process for dynamics discrimination. The binary hypotheses are $H_0$ and $H_1$ with respective prior probabilities $q_0$ and $q_1$. A known initial state evolves under hypothesis Hamiltonian $H_0$ (or $H_1$) to the final state $\rho_0$ (or $\rho_1$). Quantum measurement M is performed to decide one of the hypotheses. The coherent control strategy and noise engineering technology are considered as resources to enhance the effectiveness of decision-making. 
 (b) and (c) depict scenarios in which the two to-be-distinguished dynamics are influenced by Hamiltonian-independent noise and Hamiltonian-dependent noise, respectively.
 }
	\label{fig1}
\end{figure}

In this article, we demonstrate that the interaction between coherent evolution and noise can push the boundaries of quantum hypothesis testing beyond unitary dynamics. We explore the full potential of noise in QHT by establishing sufficient conditions for characterizing whether noisy dynamics can surpass the success probabilities achievable under noiseless dynamics.
Moreover, even when noise is initially independent and detrimental, we show that combining coherent controls with such noise can convert it into a beneficial factor, thus elevating the overall success rate in hypothesis testing. We apply our theory to discriminate quantum dynamics in spin systems under two prevalent noise models: dephasing and amplitude-damping. Our results uncover that, within specific evolution times, noisy dynamics can surpass unitary dynamics in success probability. 
Our findings signify a substantial advancement in the field of QHT and its related applications, including quantum dynamics discrimination and quantum metrology\,\cite{RN15, RN22}.

\section*{Results}

We investigate the scenario of QHT, where our objective is to distinguish between two hypotheses ($h_0$ and $h_1$) governed by Hamiltonians $H_0$ and $H_1$, respectively. These hypotheses are associated with prior probabilities $q_0$ and $q_1$. In a fixed initial state and evolution time, the process simplifies to the discrimination of final quantum states, as depicted in Figure 1(a). Subsequently, a quantum measurement is performed on the final state, and based on the measurement result, a hypothesis is selected. The success probability of this process can be expressed as $q_0 p(h_0|\rho_0) + q_1 p(h_1|\rho_1)$, where $p(h_j |\rho_j)$ represents the success probability of correctly selecting the hypothesis $h_j$ given the state $\rho_j$.
According to Holevo-Helstrom theorem\,\cite{RN6}, the success probability, given $\rho_0$ and $\rho_1$, is upper-bounded by
\begin{equation}
p_{\mathrm{succ}}\leq \frac{ 1+D_{\operatorname{tr}}(\rho_{0},\rho_{1}) }{2},
\label{eq:psuccess}
\end{equation}
where $D_{\operatorname{tr}}(\rho_{0},\rho_{1}) = \|q_0\rho_0-q_1\rho_1\|_{\operatorname{tr}} \in[0,1]$. Here
$\|\cdot\|_{tr}$ is the trace norm that equals the sum of singular values. Without loss of generality, we will assume $q_0=q_1=\frac{1}{2}$. To effectively discriminate dynamics, one can optimize the initial state, control strategies, and measurements to increase $D_{\operatorname{tr}}(\rho_{0},\rho_{1})$, thereby improving QHT performance.

In the absence of noise, the evolution is given by $\dot{\rho}=-i[H_j,\rho]$. In this case the optimal initial state for discrimination is
$|\psi_0\rangle = \frac{|\lambda_{\max}\rangle +|\lambda_{\min}\rangle}{\sqrt{2}}$, a superposition of the eigenstates of $H_1-H_0$ corresponding to the largest and smallest eigenvalues\,\cite{RN27, RN28}. With this optimal strategy, the success probability increases at a rate of $\frac{\lambda_{\max}-\lambda_{\min}}{4}$, and perfect discrimination can be achieved over sufficient time\,\cite{RN27, RN28}. 
However, when the noises are taken into consideration, 
the problem becomes much more complicated and the optimal strategies are generally unknown. For Markovian noises, the dynamics can be described by\,\cite{RN29}
\begin{eqnarray}
\aligned
\dot{\rho} &=-i[H_j,\rho]+\sum_k [L_k\rho L_k^\dagger-\frac{1}{2}\{L_k^\dagger L_k, \rho\}],
\endaligned
\label{eq:master2}
\end{eqnarray}
where $L_k$ are the Lindblad operators that describe the noisy effects and $\{x,y \} = xy + yx$ denotes an anticommutator. In the standard framework, it is generally assumed that the Lindblad operators, $\{L_k\}$, are independent of the Hamiltonian $H_j$\,\cite{RevModPhys.94.015004,zhou2018achieving,dur2014improved,nielsen2002quantum}.
Under these assumptions, common wisdom holds that noise always hampers the discrimination as it homogeneously affects both hypotheses, reducing the success rate without aiding differentiation \cite{Cooperation2023,supp}.

The assumption that noises are independent of the system's Hamiltonian does not always hold true in quantum systems. For instance, consider the decay process from a higher energy state $|e\rangle$ to a lower one $|g\rangle$. In this case, the corresponding Lindblad operator is proportional to $|g\rangle\langle e|$, reflecting an inherent dependence on the energy levels governed by the Hamiltonian. Therefore, the decay process is intrinsically correlated with the Hamiltonian. Similarly, dephasing noise, another common type of noise, often relies on the energy eigenstates because it generally affects the off-diagonal elements of the density matrix in the energy eigenstate basis. Thus when $H_0$ and $H_1$ do not commute, $[H_0, H_1] \neq 0$, the noise operators can differ due to their reliance on distinct eigenstates. Even if $H_0$ and $H_1$ initially commute and share the same eigenstates, introducing a control Hamiltonian, $H_{c}$, can alter this situation. By ensuring that $[H_0 + H_{c}, H_1 + H_{c}] \neq 0$, the eigenstates of the combined Hamiltonians will become different. It's important to note that this control mechanism is applied solely to the system, without affecting the environmental conditions, thereby illustrating how the interplay between the system's Hamiltonian and noise processes can be manipulated and potentially break the independence assumption.

In general, the noisy dynamics can be described as
\begin{eqnarray}
\aligned
\dot{\rho} &=-i[H_j,\rho]+\sum_k [L_{jk}\rho L_{jk}^\dagger-\frac{1}{2}\{L_{jk}^\dagger L_{jk}, \rho\}], j=0,1,
\endaligned
\label{eq:master3}
\end{eqnarray}
where $\{L_{jk}\}$ are the Lindblad operators under the Hamiltonian $H_j$, which can be correlated with $H_j$ and different for $j=0,1$. In such a scenario, the noise may increase the success probability of QHT as illustrated in Fig.\,\ref{fig1}. In particular, we show that with the same probe state, $|\psi_0\rangle=\frac{|\lambda_{\max}\rangle+|\lambda_{\min}\rangle}{\sqrt{2}}$, the increasing rate of the success probability under the noisy dynamics can be higher than that of the unitary dynamics for a period of time when either of the conditions holds(see Supplementary Section I for derivation~\cite{supp}),
\begin{equation}\label{eq:condition}
    \begin{aligned}
        |x_1+w_1|&> \lambda_{\max} -\lambda_{\min},\\ 
    (w_1-x_1)^2 + 4y_1^2 + 4z_1^2 &> 4 z_1 (\lambda_{\max} -\lambda_{\min}),
    \end{aligned}
\end{equation}
here $x_1=\langle \lambda_{\max}|N_1-N_0|\lambda_{\max}\rangle$, $y_1=Re \langle \lambda_{\max}|N_1-N_0|\lambda_{\min}\rangle$, $z_1=Im \langle \lambda_{\max}|N_1-N_0|\lambda_{\min}\rangle$ and $w_1=\langle \lambda_{\min}|N_1-N_0|\lambda_{\min}\rangle$, $N_j=\sum_k [L_{jk}|\psi_0\rangle\langle\psi_0| L_{jk}^\dagger-\frac{1}{2}\{L_{jk}^\dagger L_{jk}, |\psi_0\rangle\langle\psi_0|\}]$ for $j=0,1$. These conditions are determined by the Lindblad operators and the maximal and minimal eigenvectors of $H_1-H_0$, which can be directly verified. Since $|\psi_0\rangle=\frac{|\lambda_{\max}\rangle+|\lambda_{\min}\rangle}{\sqrt{2}}$ is optimal under the unitary dynamics, the noisy dynamics that satisfy either of the conditions can thus achieve a higher success probability than the maximal success probability of the unitary dynamics within certain time period. We note that for noisy dynamics,  $|\psi_0\rangle=\frac{|\lambda_{\max}\rangle+|\lambda_{\min}\rangle}{\sqrt{2}}$ might not be optimal, further optimization of the initial probe state for the noisy dynamics might yield even better results.

In our experiment, we examine a spin interacting with magnetic fields in different directions, subject to dephasing and amplitude damping noises. The dephasing noise, which can arise from the fluctuation of the magnetic field, is described by the Lindblad operator $L_{j1} =\sqrt{\kappa_1} \sigma_{\Vec{n}_j}$, where $\kappa_1$ is the dephasing rate, $\sigma_{\vec{n}_j}=\vec{n}_j\cdot\vec{\sigma}$ with $\vec{n}_j$ represents the direction of the magnetic fields. The amplitude damping noise can be described by the Lindblad operator $L_{j2} =\sqrt{\kappa_2 p} \sigma_{\Vec{n}_j}^-$ and $L_{j3} =\sqrt{\kappa_2 \left(1-p\right)} \sigma_{\Vec{n}_j}^+$, 
where $\sigma^-_{\vec{n}_j}=|g_j\rangle\langle e_j|$ is the lowering operator, $\sigma^+_{\vec{n}_j}=|e_j\rangle\langle g_j|$ is the raising operator, $|g_j\rangle$ and $|e_j\rangle$ are the ground and the excited state of $H_j=\vec{n}_j\cdot \vec{\sigma}$. The dynamics can thus be written as
\begin{equation}
\aligned
\dot{\rho}=&-i\left[ -\gamma B_j \sigma_{\vec{n}_j}/2, \rho\right]+\kappa_{1}\left(\sigma_{\vec{n}_j} \rho \sigma_{\vec{n}_j}-\rho\right)\\ &+\kappa_{2} p \left(\sigma_{\vec{n}_j}^{-} \rho \sigma_{\vec{n}_j}^{+}-\frac{1}{2}\left\{\sigma_{\vec{n}_j}^{+} \sigma_{\vec{n}_j}^{-}, \rho\right\}\right)\\ 
&+\kappa_{2}\left( 1-p \right)\left(\sigma_{\vec{n}_j}^{+} \rho \sigma_{\vec{n}_j}^{-}-\frac{1}{2}\left\{\sigma_{\vec{n}_j}^{-} \sigma_{\vec{n}_j}^{+}, \rho\right\}\right),
  \label{eq:mainmaster}
  \endaligned
\end{equation}
$p \in [0,1]$ represents the ground state population of the steady state, and in our experiment, $p \approx 0.5 $. Here, the noises are correlated with the Hamiltonian. For example, when the magnetic field is in the $xz$-plane with $\vec{n_j}=\left(\cos \theta_j, 0, \sin \theta_j \right)$, the ground and excited state of $H_j$ are $|g_j\rangle = \frac{1}{\sqrt{2-2\sin \theta_j}}\left(\begin{matrix}
    \cos \theta_j \\ 1-\sin \theta_j
  \end{matrix}\right)$,  $|e_j\rangle = \frac{1}{\sqrt{2+2\sin \theta_j}}\left(\begin{matrix}
    \cos \theta_j \\ -(1+\sin \theta_j)
  \end{matrix}\right)$, the Lindblad operators, $\sqrt{\kappa_1} \sigma_{\Vec{n}_j}$, $\sqrt{\kappa_2 p} |g_j\rangle\langle e_j|$ and $\sqrt{\kappa_2 \left(1-p\right)} |e_j\rangle\langle g_j|$, also depend on $\vec{n_j}$, which makes them correlated with the Hamiltonian. In nuclear magnetic resonance (NMR), these noisy effects are typically characterized by the longitudinal (transverse) relaxation time, $T_1(T_2)$, here $T_1=\frac{1}{\kappa_2}$ and $T_2=\frac{2}{4\kappa_1+\kappa_2}$. In the Supplementary Section III~\cite{supp}, we show that for two-level systems as long as $T_1\neq T_2$, the second condition in Eq.\,\eqref{eq:condition} is satisfied and the success probability of the noisy dynamics can exceed the limit of the unitary dynamics. 
  We also consider a general initial probe state, $a|\lambda_{\max}\rangle+b|\lambda_{\min}\rangle$, providing sufficient conditions for surpassing the unitary limit. Of practical interest, we also specify the condition for an experimentally convenient initial state $|0\rangle$.

\subsection*{Enhanced QHT via noise.}
\begin{figure}[t]  
	\makeatletter
	\def\@captype{figure}
	\makeatother
	\includegraphics[scale=0.68]{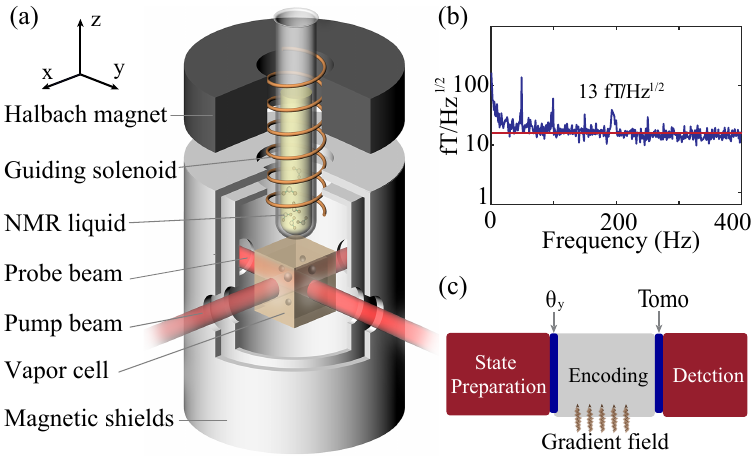} %
	\caption{\textbf{Experimental setup for quantum hypothesis testing.} (a) The proton spins in formic acid are contained in a 5\,mm NMR tube and pneumatically shuttled between a 1.3-T prepolarizing magnet and a rubidium vapor cell. A magnetic shield isolates the hypothesis testing zone from the external Earth's magnetic field. A guiding field is applied in the $z$ direction. The proton NMR signal is measured with $^{87}$Rb magnetometer. (b) Magnetic noise of $^{87}$Rb magnetometer. The noise floor is about 13\,$\rm{fT}/\rm{Hz}^{1/2}$. (c) Time sequence of quantum hypothesis testing.} 
	\label{fig2}
\end{figure}

We perform the experimental demonstration on nuclear spins using ultralow-field NMR, focusing on a two-level system represented by an uncoupled proton in $^{13}$C-formic acid (H$^{13}$COOH, where the hydroxyl proton undergoes fast chemical exchange decoupling it effectively from other spin in the molecule ). Proton spins were polarized by a 1.3-T Halbach magnet and then transferred to a shielded zone where they experienced magnetic fields $H_0$ or $H_1$, as depicted in Fig.\,\ref{fig2}(a). A guiding field was used during shuttling and discontinued upon arrival at the target region. The proton's NMR signal under $H_0/H_1$ and noise was detected with high sensitivity ($\sim$13\,$\rm{fT}/\rm{Hz}^{1/2}$ along $z$ in Fig.\,\ref{fig2}(b)) by a $^{87}$Rb magnetometer, optically pumped and probed by laser beams. Experimental details are provided in the Supplementary Section VI and VIII~\cite{supp}.

The experimental hypothesis testing process consists of three main stages (Fig.\,\ref{fig2}(c)): preparing the probe state, evolving it under the magnetic fields to be distinguished, and measuring the trace distance. The proton spins begin in a high-temperature approximated initial state,
    $\rho_{\mathrm{ini}} = \frac{\mathbbm{1}}{2} + \epsilon\frac{\sigma_z}{2}$,
where $\mathbbm{1}$ is the identity matrix and $\epsilon \approx 10^{-6}$ represents polarization. For simplicity, we set $\epsilon$ to 1 as it only scales the signal amplitude. Subsequently, the initial state evolves under either $B_0\vec{n}_0$ or $B_1\vec{n}_1$, with field directions given by $\vec{n}_j = (\cos\theta_j, 0, \sin\theta_j)$.
The evolved density matrix, $\rho_j(t)$, can be written as $\frac{\mathbbm{1} + \vec{M}_j(t)\cdot\vec{\sigma}}{2}$, where $\vec{M}_j(t)$ denotes the time-dependent proton-spin magnetization vector. The evolved proton spins' NMR signal is detected using a $^{87}$Rb atomic magnetometer. Quantum state tomography \cite{RN3} is employed to measure the three components of $\vec{M}$, thus obtaining full information about final states $\rho_0(t)$ and $\rho_1(t)$. The success probability for discriminating these states is calculated using Eq.\,\eqref{eq:psuccess}.

\begin{figure}[t]  
	\makeatletter
	\def\@captype{figure}
	\makeatother
	\includegraphics[scale=1.13]{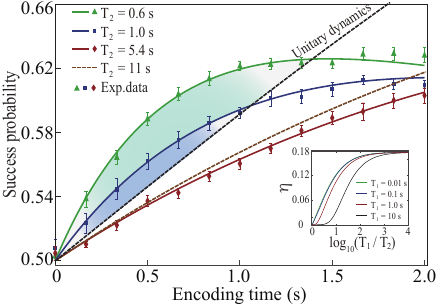} %
	\caption{\textbf{Experimental demonstration of enhanced quantum hypothesis testing.} The success probability for discriminating two magnetic fields with different directions versus the encoding time $t$. The magnetic field strength are $B_0 = B_1 = 1.86$ nT,  with $\theta_0 = 75^{\circ}$ and $\theta_1 = 30^{\circ}$. The black dashed line denotes the maximum success probability that can be achieved under the unitary dynamics. The brown dashed line represents $T_2 = 2 T_1$. The solid lines are theoretical simulations, while triangles, squares, and diamonds are experimental data under different noisy strengths.  The error bars are obtained from eight repeated experiments. In the subfigure, the maximal enhancement $\eta$ is plotted as a function of $\mathrm{log}_{10}(T_1/T_2)$.}
	\label{fig3}
\end{figure}

To investigate the impact of noises on the success probability, we conducted experiments with varying relaxation times that are simulated through the gradient field.
The experimental results are presented in Fig.\,\ref{fig3}, where the red diamond, blue square, and green triangle represent $T_2$ values of $5.4$\,s, $ 1.0$\,s, and $0.6$\,s, respectively, while $T_1$ remains constant at 5.5\,s.
When $T_2=5.4$\,s $\approx T_1$, the success probability of discriminating between the two magnetic fields falls below the limit of unitary dynamics (as described in the Supplementary Section III~\cite{supp}). This occurs because the decay rates of the nuclear-spin magnetization vectors in all directions become equal when $T_1$ is approximately equal to $T_2$. In this case, the noises become independent of the magnetic fields.
On the other hand, when $T_1$ is not close to $T_2$, indicating that the noises are dependent on the direction of the magnetic fields, the success probability exceeds the limit of unitary dynamics. This is shown as the green and blue regimes in Fig.\,\ref{fig3}. As the encoding time increases, the improvement in the success probability initially rises, reaches a maximum, and then decreases. 
Note that excessively prolonged encoding times are not beneficial as the spin-decoherence effect starts to destroy the quantum states.
Therefore, selecting an appropriate encoding time based on the noise environment is crucial to improve the accuracy of discriminating between the two Hamiltonians.

To quantify the enhancement effect, we define $\eta = \max\{ p_{\mathrm{succ1}} - p_{\mathrm{succ2}} \}$, where $p_{\mathrm{succ1}}$ and $p_{\mathrm{succ2}}$ are the success probabilities with and without noise. The inset in Fig.\,\ref{fig3} plots theoretical $\eta$ against $T_1/T_2$. As $T_1/T_2$ increases, $\eta$ grows, indicating a higher improvement in success probability due to noise. At the extreme limit of $T_1/T_2 \to \infty$, strong dephasing causes convergence to steady states defined by the dephasing dynamics.
In the strong dephasing limit, the discrimination success probability is $p_{\mathrm{succ}}=\frac{1}{2}+\frac{1}{4}|\sin(\theta_0-\theta_1)|$, with a theoretical $\eta$ of up to $\approx 17.7\%$, solely determined by the magnetic field angle difference. It is noteworthy that the inset in Fig.\,\ref{fig3} illustrates, for finite values of $T_1/T_2$, maximum enhancement $\eta$ increases as $T_1$ decreases.

\begin{figure}  
	\makeatletter
	\def\@captype{figure}
	\makeatother
	\includegraphics[scale=1.13]{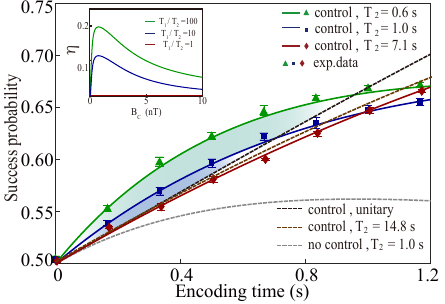} %
	\caption{\textbf{Experimental demonstration of control-assisted quantum hypothesis testing.} The experiment for the discrimination of two magnetic fields along z-direction with $B_0= 0.2$\,nT, $B_1= 2.79$\,nT, a coherent control along x-direction with $B_c= 0.75$\,nT is added. The black dashed line represents the simulated result under the unitary dynamics. The brown dashed line denotes $T_2 = 2 T_1$.
 The gray dashed line represents the simulated result without $B_c$ under the noisy dynamics with $T_2 = 1.0$\,s. The solid line is the theoretical simulation with $B_c$ under the noisy dynamics, and the experimental data are represented with triangles, squares and diamonds for different noisy strengths. The error bars are obtained from eight repeated experiments. The longitudinal relaxation time $T_1= 7.4$\,s. The inset displays the dependence of the maximum enhancement on the control field $B_c$.}
	\label{fig4}
\end{figure}

\subsection*{Control-assisted QHT.}
In scenarios where noises are initially independent of the Hamiltonian thus detrimental, it is still possible to achieve a higher success probability by leveraging the cooperation between proper coherent controls and noises. Consider a scenario where two magnetic fields are aligned along the same direction (the $z$-axis) but with different magnitudes. Without coherent controls, the Lindblad operators of the dephasing and amplitude damping noises are given by $L_1=\sqrt{\kappa_1}\sigma_z$, $L_2 = L_3^{\dagger}=\sqrt{\kappa_2}\sigma^-=\sqrt{\kappa_2} \ket{0}\bra{1}$, which are independent of the Hamiltonian. Consequently, noises can only adversely affect the success probability. However, by introducing a well-designed coherent control, we can transform the noisy operators making them correlated with the Hamiltonian. To illustrate this, we incorporate a control field, $B_c$, along the $x$-direction.
The total Hamiltonian then becomes
\begin{equation}
H_{j} =-\gamma ( B_{j}\sigma_z/2 +   B_{\rm{c}}  \sigma_x/2),
\end{equation}
where $\gamma$ is the gyromagnetic ratio of the proton. Note that without the control field, the ground and excited states of $H_j$ are $|0\rangle$ and $|1\rangle$, which are independent of $B_j$, while with the control field, the ground and excited states of $H_j$ change to $\ket{g_j}= -\sin{\frac{\theta_j}{2}}\ket{1} + \cos{\frac{\theta_j}{2}} \ket{0}$ and $\ket{e_j} = \cos{\frac{\theta_j}{2}} \ket{1} + \sin{\frac{\theta_j}{2}} \ket{0}$, here $\theta_j = \arctan{\frac{B_c}{B_j}}$.
In this case the amplitude damping noise, $L_{j2}  = L_{j3}^{\dagger} = \sqrt{\kappa_2} \sigma_{\Vec{n}_j}^-$, 
and the dephasing noise, $ L_{j1} =\sqrt{\kappa_1} \sigma_{\Vec{n}_j} $ with $\sigma_{\Vec{n}_j}= \frac{B_j \sigma_z + B_c \sigma_x}{\sqrt{B_j^2 +B_c^2}}$, are now correlated with $H_j$. 

The success probabilities with the interplay between coherent control and noise are plotted in Fig.\,\ref{fig4}, where the two magnetic fields along the z-direction are taken as $B_0 = 0.20$\,nT, $B_1=2.79$\,nT, and the control field is taken as $B_{\rm{c}}=0.75$\,nT.
The initial state is along $x$-axis. Without $B_c$, both amplitude damping and dephasing noises degrade the success probability (gray dashed line). However, when applying the control field $B_c$, the success rate can notably improve (blue and green lines), surpassing the unitary dynamics' maximum value within specific time intervals.
It is important to note that the control field does not universally convert noise into a beneficial resource. For example, when $T_1 \approx T_2$ (as indicated by the red line), the unitary limit cannot be exceeded even with a control field. 
The inset of Fig.\,\ref{fig4} presents the maximal enhancement $\eta$ achieved with varying control fields. As the magnitude of the control field increases, the enhancement initially improves but eventually starts to decline. This suggests the existence of an optimal control field that yields the highest possible enhancement. This exemplary case of QHT demonstrates that synergy of the control field and noise can lead to a boost in success probability. In the supplementary Section V, we further demonstrate that the interaction between coherent evolution and noise can positively impact the quantum Chernoff bound\cite{supp}, which serves as a measure of the asymptotic minimum error probability in distinguishing between two hypotheses, $H_0$ and $H_1$, when an ensemble of state copies, $\rho_0(t)^{\otimes n}$ and $\rho_1(t)^{\otimes n}$, is at the disposal. 

\subsection*{Discussion}
~\

The enhanced QHT scheme could be used in the field of fundamental physics research, particularly for hypothesis testing tasks that require completion within a short time frame, owing to short-lived interactions.
One specific area where this approach is valuable is in the detection of hypothetical particles beyond the Standard Model\,\cite{jiang2021search}, such as topological defect dark matter\,\cite{afach2021search} and bursts of exotic low-mass fields generated by cataclysmic astrophysical events\,\cite{dailey2021quantum}.
These exotic particle-crossing events are predicted to be transient, with durations notably shorter than the $T_2$ of the spin.
Moreover, the enhanced scheme can be effectively employed in other quantum systems including NV-center\,\cite{nv2017} and superconducting systems\,\cite{superconductor2022}.

In conclusion, our study pioneers the use of noise to enhance QHT in spin systems. We show that exploiting the dependence of noise on the Hamiltonian can boost success probability. When noise is detrimental, we demonstrate that cooperation with coherent controls can turn it into a beneficial resource. Further enhancements may be unlocked by exploring advanced strategies such as optimal quantum control\,\cite{werschnik2007quantum} and quantum error correction\,\cite{PhysRevA.100.022336}. Optimal control algorithms can optimize pulse designs for improved discrimination, while error correction can selectively suppress independent noises to increase $T_1/T_2$ ratios. Our work deepens understanding of noise's role in quantum systems and its potential to improve performance in quantum information processing tasks.

\bibliographystyle{apsrev}
\bibliography{ref2.bib}

~\

\noindent
\textbf{Data availability}.
The data that support the findings of this study are available from the corresponding authors upon reasonable request.

~\

\noindent
\textbf{Code availability}.
The codes for numerical simulation and data processing are available from the corresponding authors upon reasonable request.

~\

\noindent
\textbf{Acknowledgements}. 
This work was supported by the Innovation Program for Quantum Science and Technology (Grant No. 2021ZD0303205),
National Natural Science Foundation of China (grants nos. 11661161018, 11927811, 12004371, 12150014, 12205296, 12274395), Youth Innovation Promotion Association (Grant No. 2023474), National Key Research and Development Program of China (grant no.~2018YFA0306600) and the Research Grants Council of Hong Kong (Grant No. 14307420, 14308019,14309022).

~\

\noindent
\textbf{Author contributions}.
M.J., H.Y., and X.P. conceived the project. H.Y. and L.W. conceived the relevant theoretical constructs. Q.L. performed the measurements and analyzed the data. Z.W. assisted with the experiment and analysis.
All authors contributed to analyzing the data, discussing the results and writing the manuscript. 

~\

\noindent
\textbf{Competing interests}.
The authors declare no competing interests.

\end{document}